\documentclass[letterpaper, 10 pt, conference]{ieeeconf}  

\IEEEoverridecommandlockouts                              

\title{\LARGE \bf
Decoding Error-Related Potentials under Multisensory Feedback with Varying Congruency
}

\author{Yixin Liu$^{1}$, Kang Yin$^{1}$, Hye-Bin Shin$^{2}$, and Seong-Whan Lee$^{1}$
\thanks{*This work was partly supported by Institute of Information \& Communications Technology Planning \& Evaluation
(IITP) grant funded by the Korea government (MSIT) (No. RS-2019-II190079, Artificial Intelligence Graduate School Program (Korea University), and No. RS-2024-00336673, AI Technology for Interactive Communication of Language Impaired Individuals).}
\thanks{$^{1}$Y.-X. Liu, K. Yin, and S.-W. Lee are with the Department of Artificial Intelligence, Korea University, Anam-dong, Seongbuk-ku, Seoul 02841, Korea. \tt{\small \{yixin\_liu, charles\_kang, sw.lee\}@korea.ac.kr}}%
\thanks{$^{2}$H.-B. Shin is with the Department of Brain and Cognitive Engineering, Korea University, Anam-dong, Seongbuk-ku, Seoul 02841, Korea. \tt{\small  hb\_shin@korea.ac.kr}}%
}

\usepackage{cite}
\usepackage{amsmath,amssymb,amsfonts}
\usepackage{algorithmic}
\usepackage{graphicx}
\usepackage{textcomp}
\usepackage{xcolor}

\usepackage{booktabs}
\usepackage{multirow}

\begin{document}

\maketitle
\thispagestyle{empty}
\pagestyle{empty}

\begin{abstract}

Error-related potentials (ErrPs) are widely studied neural signatures associated with error processing in human–machine interaction. In realistic settings, error perception often occurs under heterogeneous multisensory feedback, where variability induced by sensory modality and feedback congruency poses challenges for reliable ErrP decoding. In particular, incongruent feedback is associated with increased decoding difficulty and reduced classification performance. To address this challenge, we investigate learning strategies for robust ErrP decoding under multimodal visual, auditory, and tactile feedback with controlled sensory congruency. We adopt a multi-branch EEGNet-based architecture with auxiliary supervision to improve robustness across heterogeneous conditions, without relying on explicit modality-specific assumptions. Experiments were conducted using a maze-observation task with unimodal, bimodal, and trimodal feedback configurations. Across subjects, the proposed approach achieved consistent classification performance across heterogeneous sensory conditions and showed improved accuracy compared to baseline EEGNet models, particularly under multimodal feedback. These results suggest that appropriate architectural design and training strategies can improve the stability of ErrP decoding under heterogeneous multisensory conditions.

\end{abstract}

\section{Introduction}
Error-related potentials (ErrPs) are widely studied neural signatures of human error perception in electroencephalography (EEG) signals~\cite{C5}, and have been investigated in a range of brain-computer interface (BCI) applications, including robotic control~\cite{SalazarGomez2017} and P300-based spelling systems~\cite{Combaz2012}, and other EEG-based BCI paradigms such as imagined-speech decoding~\cite{lee2023towards}, and broader BCI benchmarks and applications~\cite{jeong20222020}. Most existing ErrP studies have been conducted under controlled laboratory conditions, where sensory input is relatively simple and tightly constrained~\cite{Chavarriaga2012,Schiatti2019}. However, in realistic settings, neural responses are embedded in rich multisensory environments, in which visual, auditory, and tactile inputs often co-occur. Empirical studies in real-world interactive systems and multimodal biosignal acquisition have shown that sensory feedback and multimodal signals can substantially influence cognitive load, situational awareness, and neural decoding contexts~\cite{Wu2025,jeong2020multimodal}.

From a neural perspective, multisensory feedback introduces additional variability into EEG signals associated with error monitoring. Previous work has demonstrated that the brain dynamically integrates and reweights concurrent sensory inputs depending on their congruency and reliability~\cite{PesnotLerousseau2022,Franzen2020}, suggesting that error-related responses may exhibit increased temporal variability under multisensory conditions. Consistent with this view, electrophysiological studies have shown that congruent and incongruent multisensory feedback induces distinct EEG dynamics and temporal patterns~\cite{Ozawa2024,Zhao2023}.

From a decoding perspective, increased temporal variability and context dependence pose a fundamental challenge for conventional EEG classifiers, which may overfit to specific sensory configurations and fail to generalize across heterogeneous conditions.

Recent studies in EEG decoding and broader neural or physiological signal classification have shown that learning-based models can improve robustness and generalization under heterogeneous data conditions~\cite{ahn2022multiscale,kim2022automatic,zhao2020diagnosis}. In particular, multi-branch and related convolutional neural networks enable parallel or complementary feature extraction at different temporal scales and have been shown to be effective even in constrained EEG decoding settings~\cite{single_channel_multibranch,jeong2020eeg}. Related multi-branch designs and learning strategies have also been explored in multimodal, motor-imagery, and tactile-oriented BCI paradigms, including settings involving session-dependent variability~\cite{tsanet,lee2022motor,lee2020sessionnet}, suggesting their potential for handling sensory-induced variability. However, these approaches have primarily focused on sustained attention, steady-state responses, or task-specific decoding, rather than transient error-related potentials arising under multisensory feedback.

In realistic human-machine interaction scenarios, error perception often occurs under multisensory and potentially conflicting sensory cues, further increasing neural variability~\cite{Tessadori2017}. These considerations motivate learning strategies that jointly address temporal dispersion, cross-condition heterogeneity, and variability induced by multisensory context, while remaining applicable to passive human–machine interaction settings involving observation-based error monitoring.

From a methodological perspective, the proposed framework addresses key challenges in multisensory ErrP decoding at three complementary levels: data-level augmentation to reduce sensitivity to precise temporal alignment, architecture-level design to capture heterogeneous temporal patterns through parallel feature extraction, and training-level strategies to promote robust generalization across sensory conditions. Together, these components form a unified approach for improving decoding stability under realistic multisensory feedback.

In this study, we address the challenge of robust ErrP decoding in passive observation tasks characterized by heterogeneous sensory contexts. Rather than focusing on a single stimulus configuration, we consider a range of unimodal and multimodal conditions with varying sensory mappings, which introduce substantial variability in both the timing and expression of error-related neural responses. To mitigate this variability, we propose a pooled training framework that integrates segmentation-and-reconstruction data augmentation with a multi-branch EEGNet architecture and auxiliary modality supervision. This design aims to learn representations that are stable across sensory conditions while preserving sensitivity to error-related neural signatures.

\section{Methods}

\subsection{Problem Formulation}

Let $\mathbf{X} \in \mathbb{R}^{C \times T}$ denote an EEG epoch, where $C$ is the number of EEG channels and $T$ is the number of temporal samples within the analysis window. Each epoch is associated with a binary label $y \in \{0,1\}$, where $y=1$ indicates a participant-reported error perception (ErrP) and $y=0$ denotes a non-error trial.

The objective is to learn a parameterized classifier $f_\theta(\cdot)$ that maps an input EEG epoch to a binary error-related response label under heterogeneous sensory feedback conditions,
\begin{equation}
\hat{y} = f_\theta(\mathbf{X}).
\end{equation}
The network architecture and training strategy used to instantiate $f_\theta(\cdot)$ are described in the following subsections.

\begin{figure}[t]
    \centering
    \includegraphics[width=0.95\linewidth]{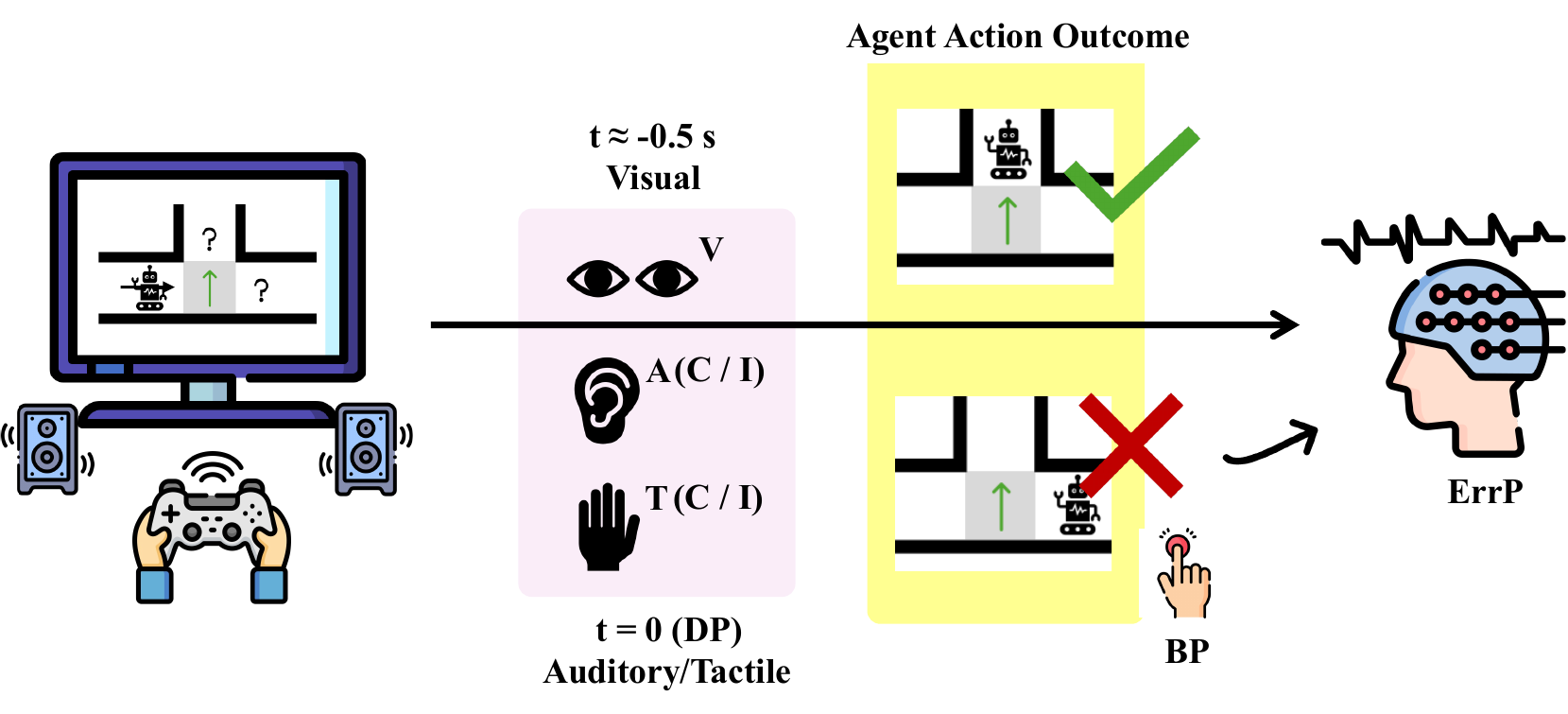}
\caption{Overview of the experimental paradigm and trial timeline. A visual cue was presented approximately 0.5\,s before the decision point (DP), whereas auditory and/or tactile feedback was delivered at the DP timestamp. Auxiliary auditory (A) and tactile (T) feedback could be congruent (C) or incongruent (I) with the visual cue--action relationship. After the agent executed a correct or erroneous action, participants pressed a button when they perceived an error, and ErrP responses were analyzed relative to the button-press (BP) event.}
    \label{fig:experimental_paradigm}
\end{figure}

\subsection{Experimental Paradigm}

The experiment was designed to elicit ErrPs associated with subjective error perception, as illustrated in Fig.~\ref{fig:experimental_paradigm}. Participants passively observed an autonomous agent navigating a two-dimensional maze environment presented on a monitor.

At predefined locations in the maze, referred to as decision points (DPs), the agent approached a junction at which a directional instruction would be provided. A decision point denotes a predefined spatial location in the maze, and the corresponding DP event timestamp in the experimental log was defined as the time when the agent reached that location. Depending on the experimental condition, additional sensory feedback could be presented in the auditory and/or tactile modalities in addition to the visual cue. Auditory feedback was delivered as brief tones, and tactile feedback was provided via vibrotactile stimulation. For each auxiliary modality, feedback was defined relative to the visual cue--action relationship, i.e., whether the executed movement was consistent with the direction indicated by the visual arrow.

Auditory feedback consisted of two tones (positive vs.\ negative), and tactile feedback consisted of two vibration intensities (weak vs.\ strong). In \emph{congruent} conditions, auxiliary feedback was aligned with the visual cue--action relationship, such that movements consistent with the visual cue were accompanied by a positive tone and/or weak vibration, whereas inconsistent movements were accompanied by a negative tone and/or strong vibration. In \emph{incongruent} conditions, this mapping was inverted for the corresponding modality.

A visual arrow cue was presented shortly before the agent reached the decision point (approximately 0.5\,s prior to the DP timestamp) to indicate the recommended movement direction, whereas auditory and/or tactile feedback, when present, was delivered at the DP timestamp. Following cue presentation, the agent executed a movement that could be either consistent or inconsistent with the indicated direction. Participants were instructed to monitor the agent’s behavior and to press a response button whenever they perceived the executed movement as inconsistent with the visual cue.

Based on these definitions, multisensory conditions were constructed by combining visual information with auditory and/or tactile feedback under congruent or incongruent mappings. Bimodal conditions included visual--auditory (VA) and visual--tactile (VT) combinations, whereas trimodal conditions combined visual, auditory, and tactile feedback (VAT). The unimodal V-only condition included only visual information without additional sensory feedback.

Condition labels are abbreviated throughout the manuscript as follows: V denotes the visual-only condition, whereas VA, VT, and VAT denote visual--auditory, visual--tactile, and visual--auditory--tactile conditions, respectively. The suffixes ``-c'' and ``-i'' indicate congruent and incongruent mappings. Thus, VA-c and VT-c denote congruent bimodal conditions, whereas VA-i and VT-i denote incongruent bimodal conditions. For trimodal conditions, VAT-c denotes the congruent trimodal condition, whereas VAT-A-i, VAT-T-i, and VAT-AT-i denote incongruency in the auditory modality, tactile modality, or both auxiliary modalities, respectively.

Each bimodal and trimodal condition was presented in one block of 36 trials and repeated twice per participant, yielding 72 trials per condition. To mitigate the relatively smaller number of trials in the unimodal configuration, the V-only condition was presented in two distinct blocks, each repeated twice, resulting in 144 trials. In total, each participant completed 720 trials across all conditions.

\subsection{Participants}

Twelve healthy adults (7 females and 5 males) participated in the experiment. Participants were in their twenties to early thirties at the time of the experiment, and all were enrolled in or had completed graduate-level education.
All participants reported normal or corrected-to-normal vision and no history of neurological or sensory disorders. Prior to the experiment, participants were informed about the
experimental procedure and provided written informed consent in
accordance with the Declaration of Helsinki.

The study protocol was reviewed and approved by the Institutional Review
Board (IRB) of Korea University. All participants completed the full
experimental session without premature termination, and no datasets
were excluded due to non-compliance or excessive artifacts.

\subsection{EEG Acquisition and Preprocessing}

EEG signals were recorded using a 64-channel actiCAP system (BrainAmp, Brain Products GmbH, Germany), with electrodes positioned according to the international 10–20 system. The data were acquired at a sampling rate of 1,000 Hz, with FPz serving as the ground and FCz as the reference during recording.

The recorded signals were first re-referenced to the common average and band-pass filtered between 1 and 30 Hz to suppress slow drifts and high-frequency noise \cite{Spuler2017}. Independent component analysis (ICA), implemented in EEGLAB \cite{Delorme2004}, was applied to the continuous EEG data to identify and remove ocular and muscular artifacts. Artifact-related components were manually rejected based on their spatial topographies and temporal characteristics.

Continuous EEG signals were then segmented into epochs based on event timestamps defined by trial type. Given that ErrPs are typically characterized by prominent frontocentral activity, the FCz channel was selected for subsequent analysis, consistent with prior studies demonstrating its sensitivity to ErrP components \cite{SorianoSegura2023,Ahkami2021}.

For trials in which participants reported an incorrect movement, epochs were time-locked to the button-press (BP) timestamp, which was used as an index of subjective error awareness \cite{Wirth2020,Vocat2008,BrainSci2024}. For trials without a button press, epochs were time-locked to the corresponding decision-point (DP) timestamp, defined as the moment when the agent reached the predefined decision location in the maze.

This asymmetric event alignment reflects the nature of the classification task, which aims to decode error-related neural responses associated with participant-reported errors in passive observation settings, where the timing of error awareness is not strictly locked to external stimulus onset but is instead driven by internal cognitive processing.

Each epoch spanned from $-0.2$~s to $0.4$~s relative to event onset. Baseline correction was applied using the pre-event interval ($-0.2$~s to $0$~s). Trials containing excessive residual artifacts were excluded following standard preprocessing criteria.

For each retained epoch, signal amplitudes were normalized using trial-wise $z$-score normalization, such that each trial had zero mean and unit variance. This normalization reduces inter-trial and inter-subject variability in overall signal amplitude, allowing the classifier to focus on temporal patterns associated with error-related neural responses. The resulting epochs were used as inputs for subsequent classification and model training.

\begin{figure*}[t]
  \centering
  \includegraphics[width=0.9\textwidth]{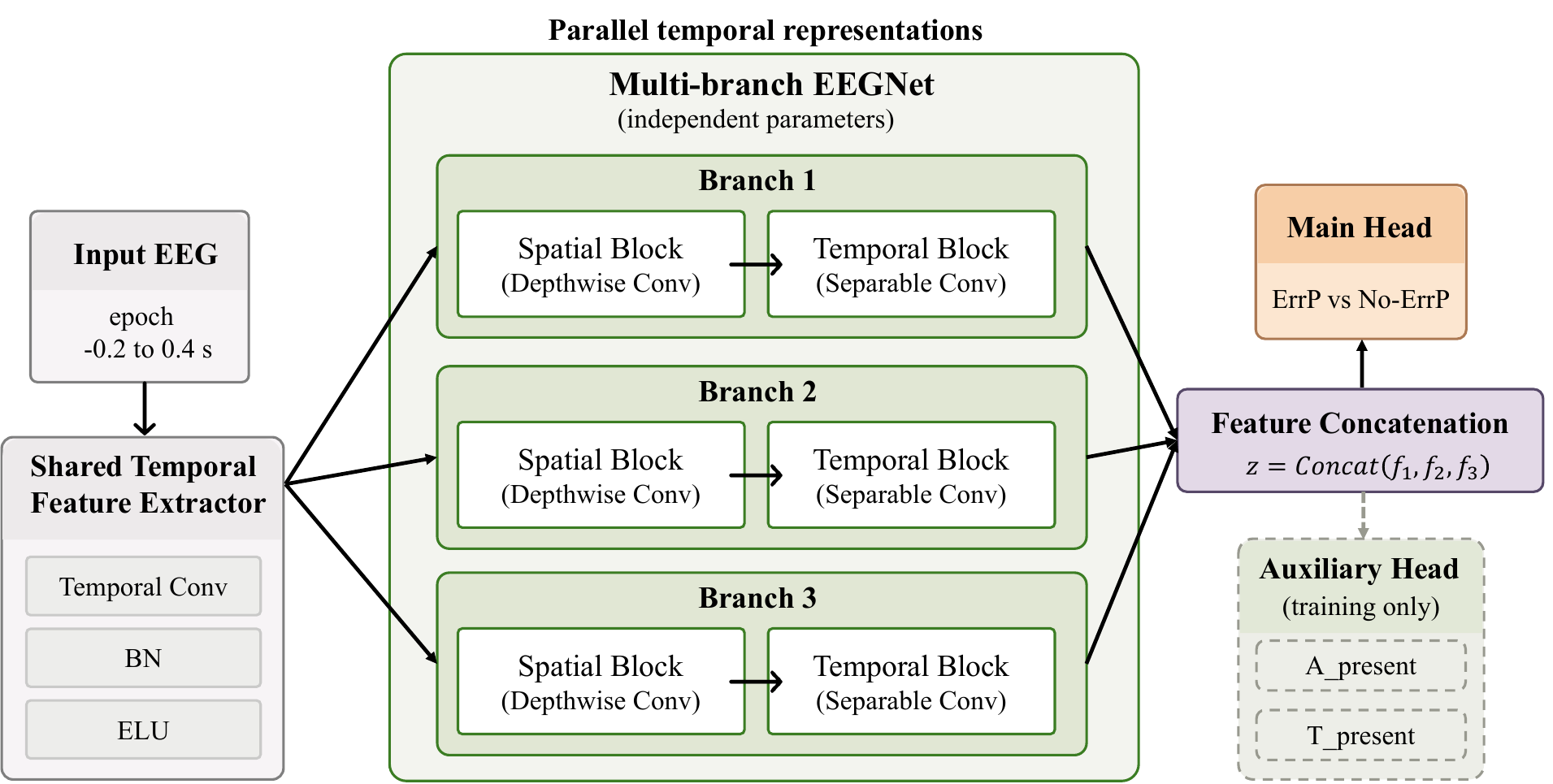}
  \caption{Architecture of the proposed multi-branch EEGNet model.
  An input EEG epoch is first processed by a shared temporal feature extractor to capture low-level temporal patterns associated with error-related neural responses.
  The extracted features are then broadcast to three parallel branches with identical architectures but independent parameters.
  Each branch consists of a spatial block followed by a temporal block, enabling the learning of complementary temporal representations without explicit modality-specific assignments.
  The outputs of the three branches are fused via feature-wise concatenation and fed into a main classification head for binary discrimination between error-related and non-error trials.
  During training, an auxiliary head is additionally attached to the fused representation to predict the presence of auditory and tactile feedback, providing regularization under pooled training across sensory conditions.
  The auxiliary head is used only during training and is removed at inference time.}
  \label{fig:multibranch_architecture}
\end{figure*}

\subsection{Model Architecture}

To classify error-related neural responses under different sensory conditions, we employed a multi-branch neural network architecture based on EEGNet~\cite{eegnet}. The proposed architecture is designed to extract complementary temporal representations from EEG signals elicited under heterogeneous sensory feedback, while maintaining a lightweight structure suitable for limited-channel settings. As illustrated in Fig.~\ref{fig:multibranch_architecture}, the model consists of a shared temporal feature extractor followed by three parallel convolutional branches, feature fusion via concatenation, and a final classification head.

The input to the network is an EEG epoch of size $C \times T$, where $C$ denotes the number of channels and $T$ the number of temporal samples. In this study, epochs were extracted from the fronto-central electrode FCz and segmented within a fixed time window relative to the behavioral event.

The shared temporal feature extractor applies a temporal convolution along the time dimension to capture low-level temporal patterns commonly associated with error-related neural responses. This block is shared across all branches to ensure that basic temporal features are learned consistently before branch-specific processing.

Following the shared block, the network splits into three parallel branches with identical architectures but independent parameters (Fig.~\ref{fig:multibranch_architecture}). Each branch is composed of two sequential stages: a spatial block and a temporal block. The spatial block employs depthwise convolution to learn spatial projections of the shared temporal features, while the temporal block adopts a separable convolution to further model branch-specific temporal dynamics. Although the branches are not explicitly assigned to modalities, consistent performance improvements across heterogeneous conditions indicate that the multi-branch design enables the model to better accommodate variability in temporal dynamics, rather than merely increasing model capacity. The number of branches was selected empirically to provide sufficient representational diversity without substantially increasing model complexity.

The outputs of the three branches are combined through feature-wise concatenation to form a unified representation,
\begin{equation}
\mathbf{z} = \mathrm{Concat}\big( f_1(\mathbf{X}),\, f_2(\mathbf{X}),\, f_3(\mathbf{X}) \big),
\end{equation}
where $f_i(\cdot)$ denotes the feature transformation implemented by the $i$-th branch. The fused feature vector $\mathbf{z}$ is then passed to a linear classification head to predict whether the observed trial corresponds to an error-related response or a non-error trial. This multi-branch design enables the model to represent diverse temporal patterns associated with multisensory processing without imposing explicit modality-specific assumptions.

\subsection{Training Procedure}

EEG classification was performed using a supervised learning framework. For condition-specific experiments, models were trained and evaluated separately for each sensory condition within each subject. For pooled experiments, trials from all sensory conditions were combined to train a single model per subject. In all settings, five-fold stratified cross-validation was applied in a trial-wise manner to ensure a balanced distribution of error and non-error trials across folds. Classification performance was reported as the mean and standard deviation of accuracy across cross-validation folds, and subject-level results were subsequently averaged across participants.

Given an EEG epoch $\mathbf{X}\in\mathbb{R}^{C\times T}$, the network outputs class logits $\mathbf{o}=f_\theta(\mathbf{X})\in\mathbb{R}^{2}$, from which class probabilities are obtained via a softmax function, $\hat{p}_{k}=\mathrm{softmax}(\mathbf{o})_{k}$ for $k\in\{0,1\}$. Model parameters $\theta$ were optimized by minimizing a class-weighted cross-entropy loss,
\begin{equation}
\mathcal{L}_{\text{main}}
= -\frac{1}{N}\sum_{i=1}^{N} w_{y_i}\log \hat{p}_{i,y_i},
\end{equation}
where $y_i\in\{0,1\}$ denotes the ground-truth label of the $i$-th trial, $\hat{p}_{i,y_i}$ is the predicted probability of the true class, and $w_{y_i}$ is the corresponding class weight. Class weights were applied to ensure stable optimization under pooled training and cross-validation, where minor class imbalance may arise across folds and augmented samples.

Under pooled training across sensory conditions, auxiliary supervision was introduced to regularize representation learning. In addition to the main error classification task, the network was jointly trained to predict the presence of auditory and tactile feedback using auxiliary classification heads attached to the fused feature representation (Fig.~\ref{fig:multibranch_architecture}). Auxiliary labels $a_i,t_i\in\{0,1\}$ were derived from the experimental condition of each trial. The total training loss was defined as
\begin{equation}
\mathcal{L}
= \mathcal{L}_{\text{main}}
+ \lambda\left(\mathcal{L}_{A}+\mathcal{L}_{T}\right),
\end{equation}
where $\mathcal{L}_{A}$ and $\mathcal{L}_{T}$ denote class-weighted cross-entropy losses for auditory and tactile modality prediction, respectively, and $\lambda$ controls the contribution of auxiliary supervision. Auxiliary heads were used only during training and were removed at inference time, such that the deployed model relies solely on the main classification head.

\begin{table*}[t]
\caption{Classification Accuracy (\%) Under Different Training Settings and Model Variants}
\label{tab:full_comparison_results}
\centering
\begin{tabular}{lcccc}
\toprule
\textbf{Condition}
& \textbf{EEGNet}
& \textbf{EEGNet+S\&R}
& \textbf{Multi-Branch+S\&R}
& \textbf{Multi-Branch+S\&R+Aux} \\
\midrule
V            & $79.23 \pm 5.81$ & $80.68 \pm 5.53$ & $80.85 \pm 6.06$ & $81.71 \pm 5.90$ \\
VA-c       & $84.40 \pm 7.61$ & $85.06 \pm 7.70$ & $86.64 \pm 4.16$ & $88.08 \pm 6.38$ \\
VA-i    & $77.39 \pm 6.56$ & $79.29 \pm 6.97$ & $82.24 \pm 5.67$ & $82.06 \pm 4.79$ \\
VT-c      & $87.98 \pm 4.29$ & $88.90 \pm 3.67$ & $90.02 \pm 4.25$ & $89.93 \pm 3.77$ \\
VT-i     & $83.09 \pm 8.50$ & $83.80 \pm 8.04$ & $86.12 \pm 7.11$ & $86.34 \pm 5.89$ \\
VAT-c      & $85.77 \pm 5.44$ & $87.50 \pm 5.42$ & $88.53 \pm 4.09$ & $91.20 \pm 4.86$ \\
VAT-A-i  & $82.18 \pm 5.46$ & $84.50 \pm 4.85$ & $84.47 \pm 5.39$ & $87.50 \pm 3.80$ \\
VAT-T-i  & $83.95 \pm 4.83$ & $85.12 \pm 5.27$ & $87.29 \pm 3.92$ & $89.58 \pm 5.54$ \\
VAT-AT-i & $80.53 \pm 8.63$ & $83.48 \pm 5.99$ & $84.43 \pm 5.34$ & $86.46 \pm 4.32$ \\
\bottomrule
\multicolumn{5}{l}{\footnotesize Values are reported as mean $\pm$ SD across subjects ($n=12$).}
\end{tabular}
\end{table*}

To improve generalization under limited training data, segmentation-and-reconstruction (S\&R) augmentation was applied to the training set within each cross-validation fold. Each EEG epoch was divided into $M$ non-overlapping temporal segments, and synthetic samples were generated by recombining segments sampled from different trials belonging to the same class:
\begin{equation}
\tilde{\mathbf{X}}
= \mathrm{Concat}\big(\mathbf{X}^{(1)}_{i_1}, \mathbf{X}^{(2)}_{i_2}, \dots, \mathbf{X}^{(M)}_{i_M}\big),
\end{equation}
where $\mathbf{X}^{(m)}_{i_m}$ denotes the $m$-th segment selected from the $i_m$-th trial. For pooled training with auxiliary supervision, recombination was further constrained to trials sharing the same error label and auxiliary modality configuration, ensuring consistency of both main and auxiliary labels. Augmentation was applied only to the training data to avoid information leakage.

All models were trained using the Adam optimizer with a fixed learning rate, which was kept constant across all experiments. Training was conducted for a fixed number of epochs, and classification performance was evaluated on the held-out test fold within each cross-validation split.

\begin{figure}[t]
    \centering
    \includegraphics[width=0.85\linewidth]{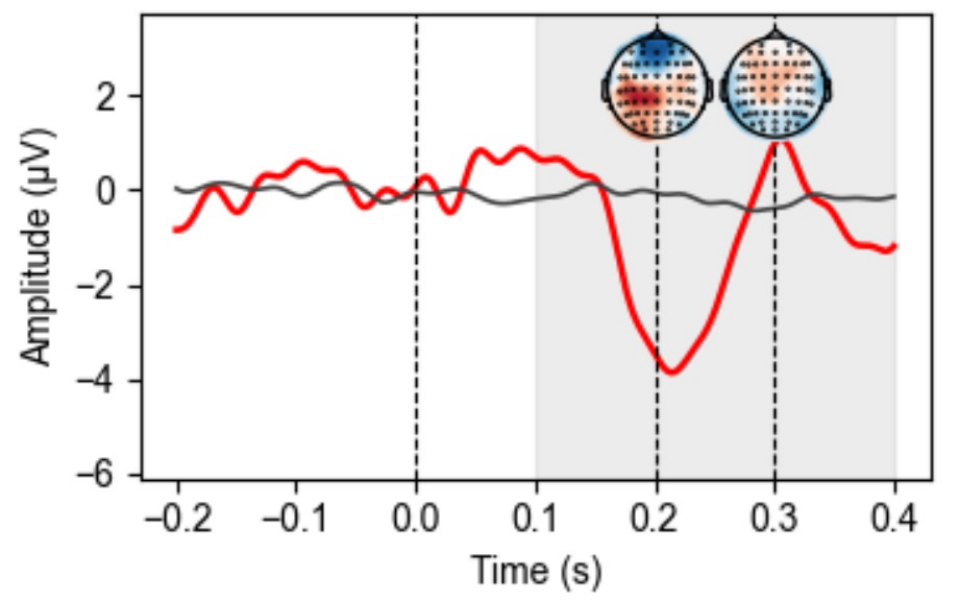}
    \caption{Grand-average EEG responses at FCz for correct trials (gray line, time-locked to DP) and error trials (red line, time-locked to button press). Topographical maps display the scalp distribution of the difference wave (error minus correct) averaged across the corresponding shaded intervals. In the topographic maps, blue denotes negative values and red denotes positive values.}
    \label{fig:cic}
\end{figure}

\section{Results}

\subsection{Grand-Average ErrP Waveforms}

To verify that the experimental paradigm reliably elicited canonical error-related potentials, we examined grand-average EEG responses at FCz. Fig.~\ref{fig:cic} shows the averaged waveforms for correct trials (time-locked to DP, without button press) and error trials (time-locked to BP, with button press).

Error trials exhibited a clear negative deflection around 200 ms (N200) followed by a positive peak near 300 ms (P300), consistent with established ErrP morphology \cite{Patel2005}. In contrast, correct trials did not show a comparable structured response. The temporal profile of these components aligns with the time window used for classification, supporting the relevance of the selected epoch range.

Although the two trial types were aligned to different event anchors, which may introduce motor-related activity in BP-locked trials, the presence of canonical ErrP components suggests that the extracted epochs captured meaningful error-related neural activity. To further assess possible motor-related contributions, we inspected the scalp distribution of the error-minus-correct response shown in Fig.~\ref{fig:cic}. The difference maps showed a predominantly fronto-central distribution around the N200/P300 intervals, consistent with typical ErrP morphology, and did not exhibit a clearly lateralized sensorimotor pattern. Nevertheless, because error trials involved a button press whereas correct trials did not, residual motor-related contributions cannot be fully excluded.

\subsection{Comparison Across Model Variants}

We evaluated classification performance across a series of model variants designed to improve decoding under heterogeneous multisensory conditions, where variability in sensory feedback poses challenges for reliable ErrP detection. The compared models included a baseline EEGNet, EEGNet with segmentation-and-reconstruction (S\&R) augmentation, a multi-branch EEGNet with S\&R, and a pooled multi-branch model with auxiliary modality supervision. In particular, we focus on how these model variants handle variability across sensory configurations and how such variability influences decoding performance.

As summarized in Table~\ref{tab:full_comparison_results}, introducing S\&R augmentation consistently improved classification accuracy over the baseline EEGNet across most sensory conditions, indicating its effectiveness as a data-level regularization strategy under limited training samples. When the multi-branch architecture was further applied, additional performance gains were observed, particularly in multimodal conditions. This suggests that parallel feature extraction enables the model to capture more diverse neural patterns associated with heterogeneous sensory feedback.

The pooled multi-branch model with auxiliary supervision achieved comparable or slightly improved overall accuracy across conditions, with the most notable gains observed in trimodal (VAT) settings, where interactions among visual, auditory, and tactile feedback are more complex.

\begin{figure}[t]
    \centering
    \includegraphics[width=\linewidth]{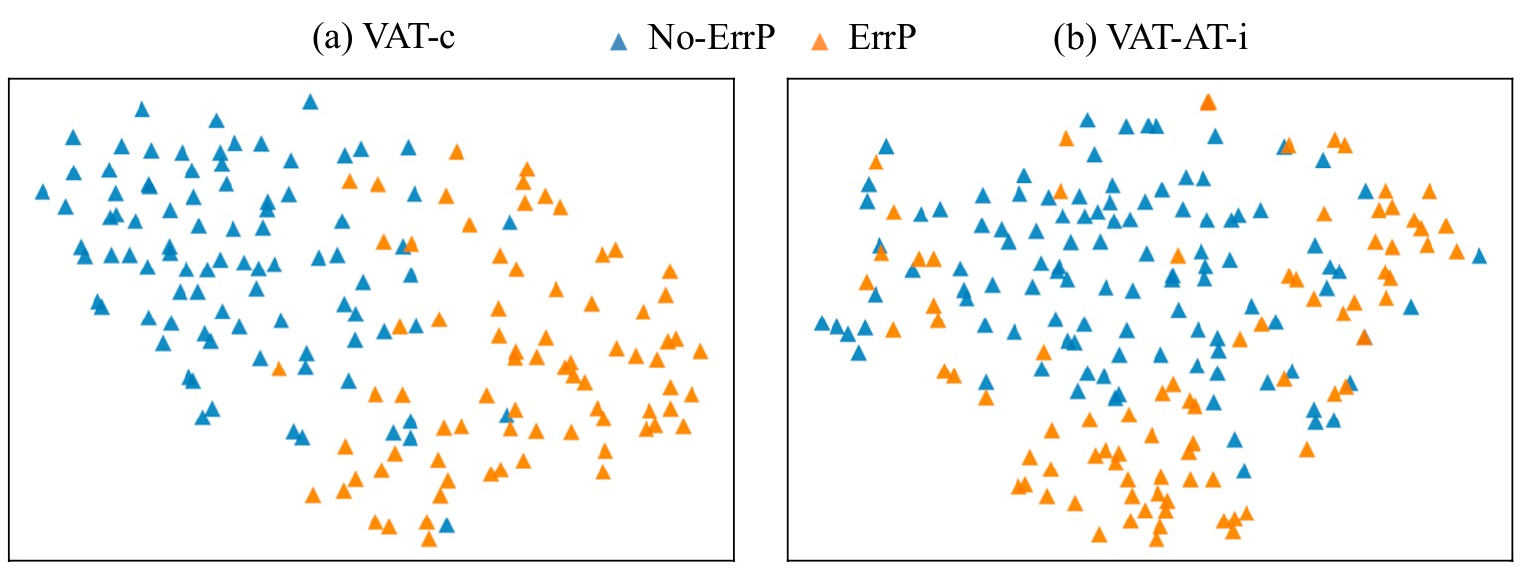}
    \caption{
    t-SNE visualization of learned feature representations from the proposed multi-branch model under two trimodal conditions. Each point corresponds to a single trial, projected from the model’s penultimate feature space after training. Blue markers denote No-ErrP trials, while orange markers denote ErrP trials.
    Left: VAT-c condition.
    Right: VAT-AT-i condition.
    For each panel, t-SNE was performed on the corresponding condition subset with identical preprocessing and hyperparameters. While both conditions exhibit partial overlap between ErrP and No-ErrP representations, the VAT-c condition shows a clearer large-scale separation structure than the VAT-AT-i condition. This qualitative difference is consistent with the quantitative classification results, where higher accuracy was observed under VAT-c feedback compared to VAT-AT-i feedback.
    }
    \label{fig:tsne_vatc_vatat_proposed}
\end{figure}

\begin{table}[t]
\caption{Paired Comparison of Classification Accuracy Between Congruent and Incongruent Conditions Under Pooled Training}
\label{tab:congruency_stats}
\centering
\begin{tabular}{lccc}
\toprule
\textbf{Comparison} & \textbf{Mean $\Delta$ Acc.} & \textbf{$p$} & \textbf{$q$} \\
\midrule
VA-c vs.\ VA-i       & +0.060 & 0.032 & 0.040 \\
VT-c vs.\ VT-i       & +0.036 & 0.027 & 0.040 \\
VAT-c vs.\ VAT-A-i   & +0.037 & 0.021 & 0.040 \\
VAT-c vs.\ VAT-T-i   & +0.016 & 0.229 & n.s. \\
VAT-c vs.\ VAT-AT-i  & +0.048 & 0.003 & 0.017 \\
\bottomrule
\multicolumn{4}{p{0.75\linewidth}}{\footnotesize
$p$: two-sided Wilcoxon signed-rank test; 
$q$: Benjamini--Hochberg FDR-adjusted value; 
n.s.: not significant ($q \ge 0.05$).
}
\end{tabular}
\end{table}

\subsection{Per-Condition Performance Under Pooled Training}

To further investigate how classification performance varies across sensory configurations, we analyzed per-condition accuracy under pooled multi-branch training with auxiliary modality supervision.

Classification performance was jointly influenced by sensory modality configuration and feedback congruency. When congruency was controlled, multimodal conditions generally exhibited higher accuracy than unimodal or bimodal settings. To quantitatively assess the effect of sensory congruency, paired statistical tests were conducted across subjects between congruent and incongruent conditions within each modality configuration. Classification accuracy was significantly higher for congruent than incongruent feedback in VA, VT, and VAT conditions after Benjamini--Hochberg FDR correction ($q<0.05$; see Table~\ref{tab:congruency_stats}), with the exception of the VAT-T-i condition, for which the difference did not reach statistical significance (two-sided Wilcoxon signed-rank tests).

Despite these variations, the pooled model maintained relatively stable performance across heterogeneous sensory conditions within subjects, indicating robust generalization under pooled training with auxiliary supervision while remaining sensitive to sensory congruency. To provide a qualitative view of how sensory congruency affects the learned representations, feature embeddings from the proposed model were visualized using t-SNE (Fig.~\ref{fig:tsne_vatc_vatat_proposed}).

\section{Discussion}

This study examined the classification of error-related EEG responses under heterogeneous multisensory feedback conditions. Across experimental settings, classification accuracy was consistently higher under congruent than incongruent feedback, indicating that conflicting sensory cues introduce additional temporal and contextual variability that complicates reliable ErrP decoding. These findings highlight sensory congruency as a key source of condition-dependent variability, consistent with the broader challenge of inter-subject variability in EEG-based BCI performance~\cite{lee2020predicting}, and emphasize the importance of decoding strategies that generalize across diverse sensory configurations rather than being tailored to a single condition.

The proposed multi-branch EEGNet showed condition-dependent performance gains, particularly in multimodal settings. Notably, the multi-branch model showed more pronounced performance gains in incongruent conditions, where sensory conflict is expected to increase temporal variability. This pattern suggests that the parallel architecture facilitates learning under heterogeneous temporal dynamics, rather than uniformly improving performance across all conditions. Such condition-dependent behavior is less consistent with a pure model-capacity effect and instead indicates that the model is better suited to capturing complex temporal structures arising under multisensory feedback.

Under pooled training across sensory conditions, auxiliary modality supervision contributed to more consistent performance across conditions, especially in more challenging incongruent and multimodal cases. Rather than primarily increasing average classification accuracy, the auxiliary loss helped preserve modality-related structure within a shared representation space, supporting more stable learning when sensory context varied across trials. These results suggest that auxiliary supervision can serve as a useful regularization mechanism in settings characterized by substantial cross-condition variability.

Several limitations should be acknowledged. First, in the current design, error trials were time-locked to the BP, whereas non-error trials were time-locked to the DP onset. Although BP-locking may introduce motor-related activity, this design reflects the subjective nature of error awareness in passive observation tasks, where the timing of error perception is not strictly aligned with external stimulus onset. Importantly, classification performance exhibited systematic differences across sensory conditions, suggesting that the model captures condition-dependent neural responses rather than relying solely on motor-related signals. Nevertheless, the contribution of motor activity cannot be entirely excluded and remains a limitation of the present study. Future work may further disentangle motor-related and error-related components by adopting consistent event anchoring or incorporating control conditions without motor responses.

Second, the present study focused on a fronto-central EEG channel (FCz), which limits the use of spatial information. This design choice reflects a common constraint in passive ErrP decoding scenarios and allows the proposed approach to be evaluated under a minimal and well-controlled input setting. Extending the framework to multi-channel EEG constitutes a natural direction for future work.

Third, error labels were derived from participant responses and therefore reflect subjective error perception rather than objective task correctness. While this choice aligns with passive human–AI interaction scenarios, it may limit direct generalization to tasks with explicit ground-truth error labels.

\section{Conclusion}

This work investigated learning strategies for ErrP decoding under multisensory feedback characterized by temporal variability and cross-condition heterogeneity. By integrating data-level augmentation, a multi-branch neural architecture, and pooled training with auxiliary modality supervision, the proposed framework achieved consistent classification performance across conditions, with more pronounced gains in multimodal and incongruent settings. These findings suggest that architectural and training-level regularization can play complementary roles in handling variability in ErrP decoding under heterogeneous sensory conditions, with implications for realistic human–machine interaction scenarios.

\addtolength{\textheight}{-12cm}   

\bibliographystyle{IEEEtran} 
\bibliography{1}

\end{document}